\documentclass[aip,pop,reprint]{revtex4-1}

\usepackage{lineno,hyperref}
\usepackage{amsmath}
\usepackage{amssymb}
\usepackage{graphicx}
\usepackage{natbib}
\usepackage[margin=1.5in]{geometry}
\modulolinenumbers[5]
\newcommand{\pd}[2]{\frac{\partial #1}{\partial #2}}
\newcommand{\gyav}[1]{\left\langle #1 \right\rangle_\mathbf{R}}

\newcommand{\dv}{\, \mathrm{d}^3 \mathbf{v}}

\begin{document}


\title{Fundamental form of the electrostatic $\delta f$-PIC algorithm \\ and discovery of a converged numerical instability}

\author{George J. Wilkie\, }
\email{wilkie@chalmers.se}
\address{University of Maryland, College Park, MD 20742}
\address{Chalmers University; 41454 Gothenburg, Sweden}

\author{William Dorland}

\address{University of Maryland, College Park, MD}

\begin{abstract}
The $\delta f$ particle-in-cell algorithm has been a useful tool in studying the physics of plasmas, particularly turbulent magnetized plasmas in the context of gyrokinetics. The reduction in noise due to not having to resolve the full distribution function indicates an efficiency advantage over standard (``full-$f$'') particle-in-cell. Despite its successes, the algorithm behaves strangely in some circumstances. In this work, we document a fully-resolved numerical instability that occurs in the simplest of multiple-species test cases: the electrostatic $\Omega_H$ mode. There is also a poorly-understood numerical instability that occurs when one is under-resolved in particle number, which may require a prohibitively large number of particles to stabilize.  Both of these are independent of the time-stepping scheme, and we conclude that they exist if the time advancement were exact. The exact analytic form of the algorithm is presented, and several schemes for mitigating these instabilities are presented.
\end{abstract}



\maketitle

\section{Introduction} \label{sec1}

Particle-in-cell (PIC) methods have been a widely used tool in plasma physics for decades. In classic ``full-$f$'' Vlasov PIC, charged particles are simulated and the fields are approximated on a grid using an appropriate interpolant. All particles of the same species are identical: the concentration of simulation particles represents the value of the distribution function at a particular location in phase space, just as it is physically. The full distribution function is solved from the Vlasov equation:
\begin{equation}\label{vlasoveqn}
\pd{f_s}{t} + \mathbf{v} \cdot \pd{f_s}{\mathbf{r}} + \frac{Z_s e}{m_s} \left( \mathbf{E} + \frac{1}{c} \mathbf{v} \times \mathbf{B} \right) \cdot \pd{f_s}{\mathbf{v}} = 0,
\end{equation}
where the distribution function for species $s$ is $f_s = f_s\left( \mathbf{r},\mathbf{v},t\right)$, with mass $m_s$, and charge number $Z_s$ (-1 for electrons). The electric and magnetic fields are $\mathbf{E}$ and $\mathbf{B}$ respectively, and are found by solving Maxwell's equation, using moments of $f_s$ to find the charge and current density. The total time derivative along particle trajectories is represented by $\mathrm{d}/\mathrm{d}t$. The PIC method is Lagrangian in the sense that a solution is obtained by the method of characteristics. Full-$f$ PIC is unweighted precisely because the right hand side of equation \eqref{vlasoveqn} is zero. This method has been well-studied and applied; its limitations are well-known because numerical dispersion relations are able to be calculated \cite{Langdon1970a,Meyers2014,Huang2015}. 

It is typical in plasma theory to expand the distribution function into a relatively constant equilibrium distribution $F_{0s}$ and a small perturbation $\delta f_s$ such that $f_s = F_{0s} + \delta f$. Aydemir \cite{Aydemir1994} took advantage of the properties of Monte Carlo integration to present a solution method which solves only for the perturbation. This method was later expanded by Parker, Lee \cite{Parker1993}, Denton, and Kotschenreuther \cite{Denton1995}, and is now known as the $\delta f$-PIC method. It greatly reduces the impact of statistical noise compared to resolving the full distribution function $f$. This scheme \emph{is} weighted in the sense that the right hand side of the kinetic equation for $\delta f$ does not vanish, so each marker caries a \emph{weight}, which changes with time along characteristics (see section \ref{sec2}). 

Because much of the dynamics has been replaced by a time-dependent weight in the $\delta f$ scheme, a numerical dispersion relation based solely on the marker trajectories is not useful. In fact, for a linear problem, the particle trajectories are entirely deterministic. Therefore, to analyze the algorithm, the changing weights play the central role; the particle trajectories only serve to complicate this analysis. Recently, work has been done\cite{Sturdevant2016} toward an analytic dispersion relation of the Vlasov $\delta f$-PIC algoirthm by approximating the distribution function as continuous rather than a collection of discrete weighted markers. This article represents an effort toward a fully analytic treatment of the gyrokinetic algorithm to seek an explanation for a numerical instability that occurs in the simplest of cases, and is converged on resolution. 



For numerical simulations, a variation of the \texttt{GSP} code \cite{Broemstrup2008} was used.

\section{The $\Omega_H$ mode} \label{sec2}

We shall concern ourselves with a simplified drift-kinetic system: that of the $\Omega_H$ mode \cite{Lee1983}. It is the simplest possible gyrokinetic system with multiple kinetic species, yet it exhibits the converged numerical instability presented here. In this regime, we make the following assumptions:
\begin{itemize}
\item Electrostatic perturbations ($\beta \rightarrow 0$)
\item Linear dynamics only (small perturbations)
\item Uniform, triply-periodic, shearless slab geometry ($\nabla \mathbf{B} = 0$)
\item Uniform Maxwellian equilibrium ($\nabla F_0 = 0$)
\item Singly-charged ions and kinetic electrons with $T_e = T_i$
\item Long-wavelength approximation ($k_\perp \rho_i \ll 1$)
\end{itemize}
Note that, as we shall discuss later, it is possible to stabilize the numerical instability by relaxing the first (electrostatic) assumption. However, the instability is still observed when any of the other listed assumptions are relaxed. These assumptions are made in this article only to simplify the analysis and to expose the basic elements that exist in a wide range of more comprehensive cases.

It will be convenient to express the gyrokinetic equation in terms of the gyroaverage of the perturbed distribution at fixed guiding center $\mathbf{R}$: 
\begin{equation}\label{gdef}
g_s \equiv \gyav{\delta f_s},
\end{equation}
so that one avoids a numerical instability resulting from multiple time derivatives \cite{Manuilskiy2000,Nielson1976}. Since the $v_\perp$ coordinate only enters the problem through the gyro-averages, and since we are taking the drift-kinetic limit, we eliminate it from the problem. Define the $\mu$-averaged distribution $\bar{g}_s$ as:
\begin{equation}\label{muavgdef}
\bar{\tilde{g}}_s \equiv 2 \pi \int\limits_0^\infty \left\langle \tilde{g}_s \right\rangle_\mathbf{r} v_\perp dv_\perp,
\end{equation}
where $\left\langle \right\rangle_\mathbf{r}$ is the gyroaverage at constant spatial position $\mathbf{r}$. In taking the long-wavelength limit, let $J_0 \approx 1$, $\Gamma_{0e} \approx 1$, and $\Gamma_{0i} \approx 1 - k_\perp^2 \rho_i^2 / 2$. 

In terms of $g_s$, under the assumptions listed above, the gyrokinetic equation\cite{Antonsen1980,Frieman1982,Abel2013} and quasineutrality read:
\begin{equation}\label{gksimple}
\pd{\bar{g}_s}{t} + v_\| \pd{\bar{g}_s}{z} = - \frac{Z_s e}{T} v_\| F_{0s\|} \pd{\phi}{z}
\end{equation}
\begin{equation}\label{qneutralitysimple}
\tilde{\phi} = \frac{2 T}{n_i e k_\perp^2 \rho_i^2} \sum\limits_s Z_s \int\limits_\infty^\infty \bar{\tilde{g}}_s dv_\|
\end{equation}
with $F_{0s\|} \equiv \left(n_{0s} / v_{ts} \sqrt{\pi} \right) e^{-v_\|^2 / v_{ts}^2}$. The distance along the straight magnetic field is $z$, with speed along the field line given by $v_\|$. Note that equations \eqref{gksimple} and \eqref{qneutralitysimple} are directly analogous to the Langmuir plasma wave in the limit $k_\perp \gg k_\|$ and $T_\| \gg T_\perp$, with an effective Debye length of $\lambda_D = n_i e^2 \rho_i / T_i$. Therefore, this instability should also be present in the Vlasov $\delta f$-PIC scheme in the appropriate limit. Note that although the form of the equations are identical, the physical interpretation of equation \eqref{qneutralitysimple} is distinct from Poisson's equation: it is instead the leading-order finite Larmor radius correction to the polarization density \cite{Dorland1993}.

We can find the corresponding dispersion relation by Laplace-transforming equation \eqref{gksimple} and inserting into equation \eqref{qneutralitysimple} to obtain:
\begin{equation}\label{omegahdisp}
k_\perp^2 \rho_i^2 = Z'\left( \frac{\omega}{k_\| v_{ti}} \right) + Z'\left( \frac{\omega}{k_\| v_{te}} \right) .
\end{equation}
The $\Omega_H$ mode is found by expanding the plasma dispersion function $Z(\zeta)$ for large argument in the dispersion relation. Including complex corrections, the frequency is approximated by:
\begin{equation}\label{omegahfreq}
\omega = \omega_0 -i \sqrt{\frac{\pi v_{ti}}{2 v_{te}}} \frac{\omega_0^3}{k_\|^2 v_{ti}^2} e^{-\omega_0^2 / k_\|^2 v_{ti}^2},
\end{equation}
where:
\begin{equation}\label{omegahfreq0}
\omega_0 \equiv \frac{k_\| v_{te}}{k_\perp \rho_i}.
\end{equation}
The wave is therefore Landau-damped. Due to its high frequency, one requires very small time-steps to resolve it, so it is easy to mistake this instability for a simple violation of the CFL condition\cite{Courant1967}. However, the numerical instability under consideration here is converged on time-step if the $\Omega_H$ mode is resolved accordingly.

\section{Matrix form of the algorithm and existence of converged numerical instability}

Before choosing a time-advancement scheme, the discrete equations that define the $\delta f$-PIC algorithm for the $\Omega_H$ mode are given in appendix \ref{derivationsec}. 

By combining equations \eqref{gkcharacteristics}, \eqref{charcurves}, \eqref{fieldgrid2particle}, \eqref{fieldfourier}, \eqref{phifromq}, and \eqref{chargegrid} , we find that the ODE for the marker weights is coupled linearly to the weights of all other markers. That is:
\begin{equation}\label{matrixeqn}
\dot{w}_\alpha = \sum\limits_\beta M_{\alpha \beta}(t) w_\beta,
\end{equation}
where:
\begin{widetext}
\begin{align}
M_{\alpha \beta}(t) &= \frac{- 2 Z_\alpha Z_\beta v_\alpha}{N_p} \sum\limits_\mathbf{k, i, j} S_3\left( \mathbf{R}_\alpha - \mathbf{r}_i \right) S_3\left( \mathbf{R}_\beta - \mathbf{r}_j \right) e^{i \mathbf{k} \cdot \left( \mathbf{r}_j - \mathbf{r}_i\right)} \frac{i k_\|}{k_\perp^2 \rho_i^2} \nonumber  \\
 &\approx \frac{- 2 Z_\alpha Z_\beta v_\alpha}{N_p} \sum\limits_\mathbf{k} |S_\mathbf{k}|^2 e^{i \mathbf{k} \cdot \left( \mathbf{R}_\beta - \mathbf{R}_\alpha \right)} \frac{i k_\|}{k_\perp^2 \rho_i^2} \label{matrixelements}
\end{align}
\end{widetext}
The last line is only an approximation because one cannot rigorously shift a discrete Fourier transform continuously. It is our experience, however, that using this approximation makes little qualitative difference to the behavior of the algorithm, and this is the analytic form taken in reference \cite{Nevins2005}. We will, however, be using the exact form unless stated otherwise. The time-dependence of the matrix elements \eqref{matrixelements} comes from that of the marker positions, through equation \eqref{charcurves}. Therefore, the $\delta f$-PIC algorithm is fundamentally a large coupled system of first-order ODEs with variable coefficients. 

One can solve equation \eqref{matrixeqn} semi-analytically only in the case where there is a single-mode (with a given $\mathbf{k_\perp}$ and $k_\|$) present, and the approximation made in equation \eqref{matrixelements} is valid. In this case, the solution is:
\begin{equation}\label{semianalyticsoln}
\mathbf{w}(t) = \mathbf{w}(t=0) \exp\left[ \mathbb{A}(t)\right];
\end{equation}
here $\exp$ is the matrix-exponential and the elements of $\mathbb{A}$ are given by:
\begin{align} \label{Adef}
A_{\alpha \beta} (t) =& \int\limits_0^t M_{\alpha \beta}(t') dt'  \\
=& M_{\alpha \beta} \frac{-i}{k_\| \left( v_{\| \beta} - v_{\| \alpha} \right)} e^{i k_\| \left( z_{\beta 0} - z_{\alpha 0} \right)}  \nonumber \\   
&\times \left( e^{i k_\| \left( v_{\| \beta} - v_{\| \alpha} \right) t } - 1 \right). 
\end{align}
When multiple Fourier-modes are allowed, this solution is not valid because then the matrix $\mathbb{M}$ fails to commute at different times ($\mathbb{M}(t_1) \mathbb{M}(t_2) \neq \mathbb{M}(t_2) \mathbb{M}(t_1)$).

\begin{figure}
\center\includegraphics[width=0.5\textwidth]{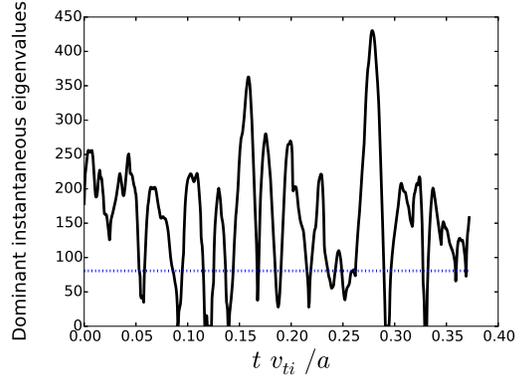}
\caption{\label{eigVSt} Time-evolution of the largest real eigenvalues of matrix \eqref{matrixelements}. The horizontal dotted line marks the approximate average growth rate of the code ($\gamma \approx 80.6$) for the parameters: $N_y \times N_z = 4 \times 4$, $L_z = 2 \pi a$, $L_y = 40 \pi \rho_i $, $N_p = 320$.}
\end{figure}

In general, for multiple modes, equation \eqref{matrixeqn} offers no immediate analytic solution. However, at any given moment during a simulation, we can take the matrix to be approximately constant, and use that to calculate the instantaneous eigenvalue spectrum. The time-evolution of the most unstable eigenvalue is given in figure \ref{eigVSt}.

\begin{figure}
\center\includegraphics[width=0.5\textwidth]{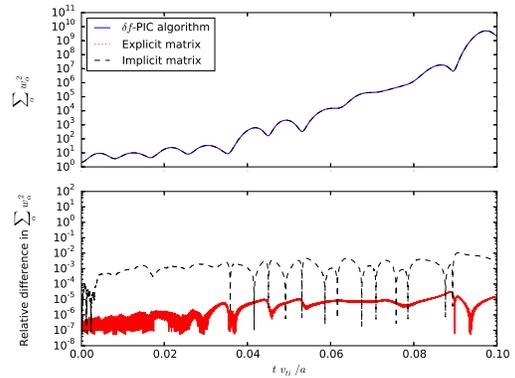}
\caption{\label{impexcomp} Comparing the $\delta f$-PIC algorithm to a direct solve of the matrix system \eqref{matrixeqn}. The sum of the squared particle weights is displayed in the upper chart. The relative difference compared to the results of an actual $\delta f$-PIC code: $| \left[\sum_\alpha w_\alpha^2\right]_\mathrm{PIC} - \left[ \sum_\alpha w_\alpha^2 \right]_\mathrm{matrix}| / \left[ \sum_\alpha w_\alpha^2\right]_\mathrm{PIC}$ is shown below.  The explicit matrix is accurate to machine precision, while the implicit scheme suffers from a small amount of numerical dissipation due to the finite-time-step. The explicit and implicit schemes used here are forward and backward Euler respectively, with $\delta t = 10^{-6} a /v_{ti}$. The resolution is $N_y = 4$, $N_z= 4$, and 20 particles per species per grid cell. A low resolution is necessary due to the need to invert a dense matrix of size $N_p \times N_p$ every time-step.} 
\end{figure}

The linear system \eqref{matrixeqn}  can even be solved implicitly:
\begin{equation}\label{implicitsystem}
\mathbf{w}(t+\Delta t) = \left(  \mathbb{M}(t+\Delta t) \Delta t - \mathbb{I} \right)^{-1} \mathbf{w}(t),
\end{equation}
where $\mathbf{w}$ is a column-vector of all the particle weights. Figure \ref{impexcomp} compares the relative difference in the solution obtained by the code \texttt{dk2d}, and solving the linear system \eqref{matrixeqn} explicitly and implicitly. Note that the explicit matrix solution is identical to within machine precision to the $\delta f$-PIC algorithm. In fact, it should \emph{be} the algorithm, with no approximations made. 

\begin{figure}
\center\includegraphics[width=0.5\textwidth]{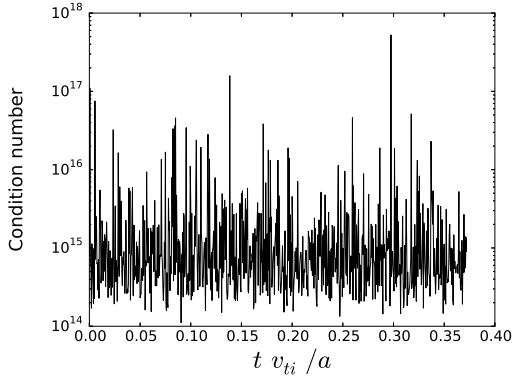}
\caption{\label{condVSt} Time-evolution of the condition number of matrix \eqref{matrixelements}. Same case as figure \ref{eigVSt}} 
\end{figure}

Another property of the matrix $\mathbb{M}$ is that it is poorly conditioned (see the evolution of the condition number in figure \ref{condVSt}). This implies that the dynamics are not reversible without a roundoff-error catastrophe, but it is unclear if this provides insight to the numerical instability discussed below.

Note that solving this matrix system is extremely inefficient compared to the particle algorithm. In a $\delta f$ particle-in-cell code, this matrix never needs to be stored, calculated, or inverted in its entirety. Nevertheless, equation \eqref{matrixeqn} is the exact analytic form of the linear algorithm. We find that the numerical instability presented in the following section is a property of this matrix system itself, and would exist even if the time advancement were exact.

\section{Characterization of the numerical instability} \label{sec4}

Here, we detail the properties of the discovered numerical instability. In what follows, it will be useful to distinguish between the two separate instabilities observed: a \emph{finite-particle} instability, which is difficult to characterize, and whose average growth rate generally decreases with increasing particle number. Indeed, there is not a clean exponential behavior associated with this numerical instability (see, \emph{e.g.}, figure \ref{npconverge}). As the number of particles increases, one ultimately finds a \emph{converged} numerical instability at some mode numbers, which does, in contrast, exhibit clear exponential/oscillatory behavior. It is this \emph{unconditional} instability that we will chiefly concerns ourselves with in this section.

The standard, minimally-resolved case in which one can observe the converged numerical instability is: $L_y = 20 \pi \rho_i$, $L_z = \pi a$, $N_y = 4$, $N_z = 4$, $m_i / m_e = 1849$, $N_c = 8000$.

\subsection{Convergence in time-step}

\begin{figure}
\center\includegraphics[width=0.5\textwidth]{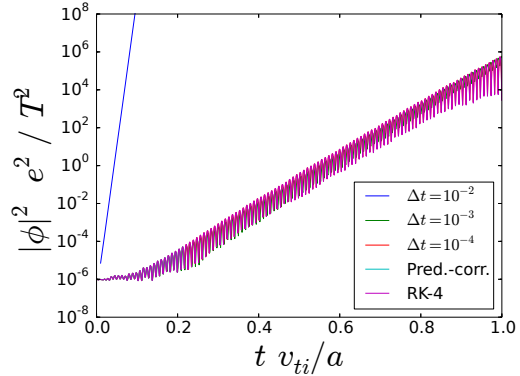}
\caption{\label{dtconverge} Demonstrating the convergence of instability growth rate on time-step size and method. Second-order Runge-Kutta is the default, with a simple predictor-corrector scheme and fourth-order Runge-Kutta also shown (the latter two have $\Delta t = 10^{-4}$). At high step-size, the simulation is wildly unstable, which is to be expected from a violation of the CFL condition. }
\end{figure}

As a basic check, we verify that we are not violating a CFL condition. By decreasing the time step and changing time-integration methods, we converge upon the same unstable solution. We are confident that we are converging upon the exact solution of the time-continuous equations (see figure \ref{dtconverge}).

\subsection{Convergence in particle number}

\begin{figure}
\center\includegraphics[width=0.5\textwidth]{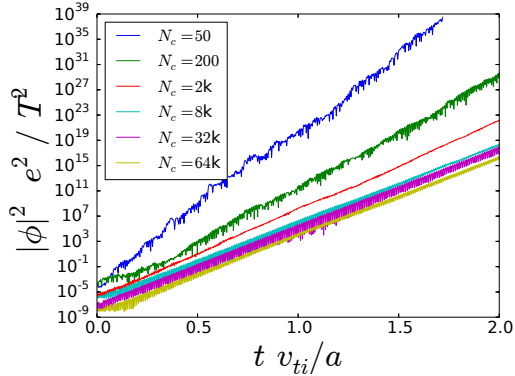}
\caption{\label{npconverge} Demonstrating the convergence of instability growth rate on particle number. Under-resolved cases ($N_c < 8000$) suffer from the poorly-behaved finite-particle instability. }
\end{figure}

In figure \ref{npconverge}, we illustrate the convergence of the unphysical growth rate on particle number. When initializing with random noise, it is expected that the initial size of the perturbation decreases with increasing particle number. However, there is a fixed growth rate one reaches at which we consider the solution converged.

Note also the non-exponential behavior when $N_c$ is below the threshold. This is the finite-particle instability, which proves troublesome in some circumstances, requiring a large number of particles per cell to stabilize. While it is expected that under-resolving the number of particles would cause a loss of accuracy, it is not clear why a numerical instability would result. It is this instability that is observed in figure \ref{npconverge} and confirmed with the $\delta f$-PIC matrix \eqref{matrixelements} to be a fundamental feature of the algorithm. It is not clear if the converged numerical instability is another aspect of this finite-particle effect, or if they are in fact two separate instabilities arising in independent circumstances.

\subsection{Scaling with parallel wave number}

\begin{figure}
\center\includegraphics[width=0.4\textwidth]{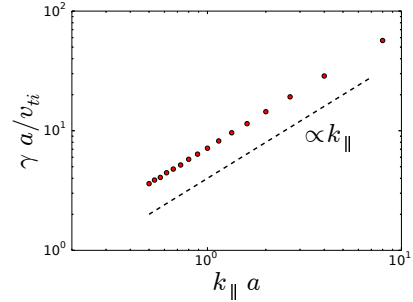}
\caption{\label{kzscaling} Dependence of the unphysical growth rate on parallel wave number. The actual power law fit is $k_\|^{0.9976}$.}
\end{figure}

When multiple modes are present (particularly a parallel mode and its counter-propagating partner) there is a clear direct linear relationship between the parallel wavenumber and the growth rate: see figure \ref{kzscaling}. This is not particularly surprising: the arbitrary parallel length scale $a$ only enters in defining $k_\parallel$ and the characteristic time $a / v_{ti}$. 


\subsection{Scaling with mass ratio}

\begin{figure}
\center\includegraphics[width=0.4\textwidth]{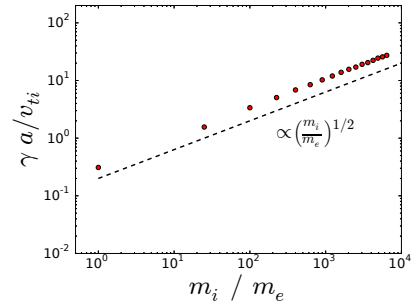}
\caption{\label{mescaling} Dependence of the unphysical growth rate on electron-ion mass ratio. Note the existence of instability at $m_e = m_i$. Actual power law fit is $\left(m_i / m_e \right)^{0.5131}$}
\end{figure}

Two kinetic species are required to observe the numerical instability. Typically, these are taken to be light electrons and heavy ions. In fact, this was used in the simplification that $\Gamma_{0e} = 1$, and $\Gamma_{0i} = 1 - k_\perp^2 \rho_i^2 /2$. As the mass ratio is adjusted using the same simplified field equation \eqref{qneutralitysimple}, we find that the growth rate scales linearly with $v_{te} / v_{ti} = \sqrt{m_i / m_e}$ (see figure \ref{mescaling}). 

Note that a positive unphysical growth rate remains even when the mass ratio is taken to be unity (as in a positronic plasma). In this limit, the approximation used in equation \eqref{qneutralitysimple} breaks down and electron Larmor radius effects play as much of a role as the ``ions''. So while the instability still exists with full Larmor radius effects, the important point here is that the separation of scales between the characteristic velocities of ions and electrons is \emph{not} responsible for the instability, although it does have a scaling effect on the growth rate.

\subsection{Scaling with perpendicular wave number}

\begin{figure}
\center\includegraphics[width=0.4\textwidth]{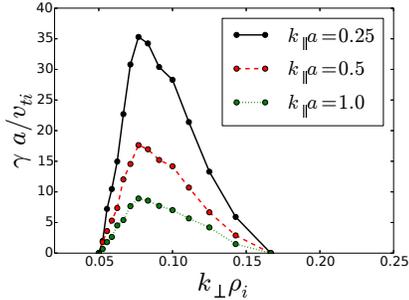}
\caption{\label{kperpscaling} Illustrating the dependence of the unphysical growth rate on perpendicular wave number at several values of $k_\parallel$.}
\end{figure}

It is found that the numerical instability is only converged for a range of wavenumbers $k_\perp$. In figure \ref{kperpscaling} is the relationship of the growth rate with $k_\perp$. A peak growth rate is observed around $k_\perp \rho_i \approx 0.08$. The threshold for stability does not change with $k_\parallel$, which is to be expected from the linear scaling. The finite-particle instability has an even more complicated dependence on $k_\perp$, which is not detailed here.
\section{Mitigation schemes} \label{sec5}

That such a catastrophic instability is fundamental to the $\delta f$-PIC algorithm is surprising since there are several examples of it reproducing good physics \cite{Chen2003,Chen2009,Mishchenko2014}. To get good results, other groups must have either avoided this particular range in parameter space, or employed one of several mitigation methods that have been found to stabilize the resolved instability. From most to least physically satisfying, this section presents possible methods of mitigating the numerical instability.

\subsection{Small, finite-$\beta$}

The kinetic Alfv\'en wave becomes the $\Omega_H$ mode in the limit $\beta \ll m_e / m_i$. As long as one avoids this regime by running at a small, but finite $\beta$, the Alfv\'en wave appears to be stable in the $\delta f$-PIC algorithm \cite{Chen2009}.

In physics, one rarely encounters plasmas of interest with $\beta < m_e / m_i$, so this is a somewhat physically-satisfying solution. Some authors have discovered modifications to the algorithm, such as a split-weight scheme \cite{Manuilskiy2000} or by using canonical momentum coordinates \cite{Mishchenko2014}, that may make the electromagnetic algorithm more efficient or accurate.

\subsection{Piecewise-constant fields}

Another way to stabilize the converged instability is to alter the way the gradient of the potential is calculated at the location of the particle. In (\ref{fieldgrid2particle}), we calculate the gradient on the grid, then interpolate that to the particles. Instead, one can use the \emph{local} gradient at the particle location given $\phi$ on the nearby grid points. Then, all particles within a grid cell would experience the same electric field. For example, if instead of (\ref{fieldgrid2particle}), we have:
\begin{equation} \label{fieldalt}
\pd{\phi}{z_\alpha} = \frac{1}{\Delta z} \left( \phi_{z_{i'+1}} - \phi_{z_{i'}} \right)
\end{equation}
\begin{equation}
i' \equiv \mathrm{mod}\left[ \mathrm{floor} \left( \frac{z_\alpha}{\Delta z} \right), N_z \right],
\end{equation}
then the algorithm appears to be only unstable to the finite-particle instability. The reason this occurs is unclear.

This method is not without its costs, however. Besides a general loss of accuracy by taking the electric field as piece-wise constant, one introduces a self-force from a particle experiencing a field from its own charge. This occurs because the interpolation from the particle to the grid is no longer symmetric with the corresponding interpolation from the grid to the particle \cite{Langdon1970a}. This can be seen by noting that the elements $M_{\alpha \alpha}$ in (\ref{matrixelements}) vanish. By altering the algorithm with (\ref{fieldalt}), this feature is lost. Though no detailed investigation on the consequences of this scheme has been performed here, this would surely introduce undesirable effects.

\subsection{Collision operator / coarse-graining}

Until now, we have considered only the collisionless problem. Implementing a physically-rigorous collision operator into $\delta f$-PIC is a challenge, and one to which the community has not yet reached consensus on an acceptable solution. The implementation of collisions used in \texttt{GSP} is based on the pitch-angle operator of Broemstrup \cite{Broemstrup2008}, which itself is an extension of the Chen-Parker coarse-graining method \cite{Chen2007a}. 

Based on our simulations, an effective pitch-angle collision frequency of about $\nu \sim 30 v_{ti} / a$ is required the stabilize the instability. For this case, the growth rate is comparable, so our interpretation here is that this strong of a collision frequency simply introduces strong enough damping to counteract the instability. 

\subsection{Mode filtering}

Perhaps the least satisfying way to stabilize the algorithm is to simply ignore the modes that are unphysically unstable. Doing so involves employing a mask in Fourier space such that after $\phi_\mathbf{k}$ is calculated, one negates a set of modes that suffer from numerical instability. This works because, as evident from figure \ref{kperpscaling}, only a range of $k_\perp$ are unstable. Although it has not been analyzed here, it is possible that in toroidal geometry with magnetic shear, a smaller range of modes might be unstable and could make this technique more palpable. Furthermore, if there is a way to \emph{a priori} predict when such modes will be unstable, filtering could be a satisfying solution if it were a function of grid resolution, which we maintain may be a possibility.


\section{Conclusion} \label{sec6}

In this work, we presented the analytic form of the $\delta f$-PIC algorithm. In doing so, a fundamental numerical instability is exposed: one that is due to under-resolution in particle number. While having too few particles certainly leads to inaccuracy \cite{Nevins2005}, there is no clear reason why a strong numerical instability should result. Furthermore, there is an even more troubling instability that is converged on particle number. There is therefore no way to use the algorithm to distinguish the instability as unphysical, and it presents itself clearly in the simplest multi-species gyrokinetic situation: the $\Omega_H$ mode in a periodic slab. 


Responsible computational physics requires a fundamental understanding of the discrete equations of a proposed algorithm and a clear expression of its limitations. There is much work to be done in this regard with the $\delta f$-PIC algorithm, and this work in addition to the approach of reference \cite{Sturdevant2016} is a good start toward this goal. A statistical analysis of the set of ODEs (\ref{matrixeqn}) is warranted, though it is not clear how such an analysis would proceed. Once a fundamental cause for the unphysical instability is found, work can proceed in mitigating the undesired behavior at a fundamental level. 

The authors would like to thank Greg Hammett, Yang Chen, Scott Parker, Ralf Kleiber, Alberto Bottino, and Benjamin Sturdevant for helpful discussions and correspondence. This material is based upon work supported by the U.S. Department of Energy, Office of Science, Office of Fusion Energy Science, under award numbers DE-FG02-93ER54197 and DE-FC02-08ER54964.
 
\appendix 

\section{Analytic form of the $\delta f$-PIC algorithm} \label{derivationsec}

This appendix details the $\delta f$-PIC method of solving equations \eqref{gksimple} and \eqref{qneutralitysimple}, taking advantage of two powerful multi-dimensional techniques: the method of characteristics, and Monte-Carlo integration.

With the method of characteristics, we can reduce an $n$-dimensional partial differential equation to a set of ordinary differential equations along characteristic curves that define the proper time derivative. This will give us the solution along any appropriate characteristic curve, headed by a \emph{marker} (or ``particle''), labelled in this work by Greek indices. The species index will be taken to be implicit in the marker index, so where convenient we will write, for example, $Z_\alpha$ instead of $Z_{s(\alpha)}$. Define a marker weight, which is just the normalized solution of the gyrokinetic equation along its characteristic trajectory:
\begin{equation}\label{wgtdef}
w_\alpha \equiv \frac{\bar{g}_{s(\alpha)} \left( \mathbf{R}_\alpha, v_{\|\alpha} \right)}{F_{0s(\alpha)}} = \frac{\bar{g}_\alpha}{F_{0\alpha}}.
\end{equation}
In terms of $w_\alpha$, equation \eqref{gksimple} becomes:
\begin{equation}\label{gkcharacteristics}
\frac{d w_\alpha}{d t} = - \frac{Z_\alpha e}{T_\alpha} v_\| \left(\pd{\phi}{z} \right)_{\mathbf{r}=\mathbf{R}_\alpha},
\end{equation}
which is the solution of the gyrokinetic equation along characteristic curves defined by:
\begin{equation} \label{charcurves}
\frac{d x_\alpha}{d t} = 0, \, \frac{d y_\alpha}{d t} = 0, \, \frac{d z_\alpha}{d t} = v_{\| \alpha}.
\end{equation}
A marker's position at any time is:
\begin{equation}\label{Rvecdef}
\mathbf{R}_\alpha (t) = \left( x_{\alpha 0} , \, y_{\alpha 0}, \, z_{\alpha 0} + v_{\| \alpha} t \right).
\end{equation}
Some authors choose to normalize the weight by the full distribution $f$ (such that $w = \delta f / f$). This would introduce a factor of $1/(1 - w_\alpha)$ to the right-hand side of (\ref{gkcharacteristics}). Even so, the numerical instability remains, and when linearized for small perturbations, \eqref{gkcharacteristics} is recovered. 

In order to solve equation \eqref{wgtdef}, it remains to find $\partial \phi / \partial z$ at the marker location $z_\alpha$. We will use a spatial grid to aid in this, with a 3D interpolant function $S_3\left(\mathbf{r}\right) = S(x / \Delta x) S(y / \Delta y) S(z / \Delta z)$, where $S$ can be one of many possible shape functions (see reference \cite{Langdon1970a}), and $\Delta x$, $\Delta y$, and $\Delta z$ are the grid spacings in the $x$, $y$, and $z$ directions respectively. Without a loss of generality, we will take $S$ here to be the linear interpolant function: 
\begin{equation}\label{Sdef}
S(x) \equiv 
\begin{cases}
1 - |x|, & \mathrm{if} \, |x| < 1 \\
0, & \mathrm{otherwise}.
\end{cases}
\end{equation}
Therefore, if we know $\partial \phi / \partial z$ on grid points labelled by $\mathbf{r}_i$, we can find the corresponding value at location of marker $\alpha$ by:
\begin{equation}\label{fieldgrid2particle}
\pd{\phi}{z_\alpha} =  \left(\pd{\phi}{z} \right)_{\mathbf{r}=\mathbf{R}_\alpha} = \sum\limits_i S_3\left(\mathbf{R}_\alpha - \mathbf{r}_i \right) \pd{\phi}{z_i}.
\end{equation}

Define the discrete Fourier transforms of an arbitrary scalar $Q\left(\mathbf{r}\right)$:
\begin{equation}\label{DFTdef}
Q_\mathbf{k} = \mathcal{F}_D\left[ Q(\mathbf{r}_i) \right] = \sum\limits_i e^{-i \mathbf{k}\cdot \mathbf{r}_i} Q(\mathbf{r}_i),
\end{equation}
\begin{equation}\label{DFTinvdef}
Q_i = \mathcal{F}_D^{-1}\left[ Q_\mathbf{k}) \right] = \frac{1}{N_g} \sum\limits_\mathbf{k} e^{i \mathbf{k}\cdot \mathbf{r}_i}\tilde{Q}(\mathbf{k}),
\end{equation}
where $N_g = N_x N_y N_z$ is the total number of grid points and $\mathbf{k}$ is discretely-valued. Now, find $\partial \phi / \partial z$ on the grid point $\mathbf{r}_i$ in terms of the Fourier modes of $\phi$:
\begin{equation}\label{fieldfourier}
\pd{\phi}{z_i} = \frac{1}{N_g} \sum\limits_\mathbf{k} e^{i \mathbf{k} \cdot \mathbf{r}_i} i k_\| \phi_\mathbf{k}.
\end{equation}

To proceed, we need a way of estimating the integral in equation \eqref{qneutralitysimple} on the grid. We do this by means of Monte-Carlo integration \cite{Aydemir1994}, using the same interpolant as in equation \eqref{fieldgrid2particle} for the spatial dependence. Monte-Carlo is a method by which one expresses an arbitrary integral (in this case a one-dimensional integral in $v_\|$) as an expectation value over some probability distribution $p$, then estimates this expectation value as a discretely-sampled average. Formally,
\begin{equation}\label{MCdef}
\int\limits_{-\infty}^{\infty} \bar{g}_s (v_\|) dv_\| = \int\limits_{-\infty}^{\infty} \frac{\bar{g}_s (v_\|)}{p(v_\|)} p(v_\|) dv_\| = \left\langle \frac{g}{p} \right\rangle_p,
\end{equation}
where $\left\langle \right\rangle_p$ is the expectation value over the probability distribution $p$, which obeys the following properties:
\begin{equation}
p > 0 \, \hspace{4mm} \forall v_\|
\end{equation}
\begin{equation}
\int\limits_{-\infty}^\infty p(v_\|) dv_\|  = 1.
\end{equation}
If we take $p = F_{0\|s} / n_{0s}$ by distributing markers according to a Maxwellian in $v_\|$, and have $N$ discrete samples of $\bar{g}_s$, then we can write:
\begin{equation}\label{MCvel}
\left\langle \frac{g}{p} \right\rangle_p = \frac{n_{0s}}{N} \sum\limits_{\alpha=1}^N w_\alpha + \mathcal{O}\left( \frac{n_{0s} \mathrm{var}\left(w\right)}{\sqrt{N}} \right). 
\end{equation}
The extension to multiple dimensions can be found in reference~\citep{Aydemir1994}. 

To account for the spatial-dependence of equation \eqref{MCvel}, we use the interpolant function $S_3$ since the location at which we want the integral (on a grid point $\mathbf{r}_i$) is in general not the same as the marker positions $\mathbf{R}_\alpha$. This is perhaps the only non-rigorous part of the algorithm and may be what is responsible for unphysical behavior at large particle number. 

Moving forward with this caveat in mind, we can estimate the charge density at the spatial grid location $\mathbf{r}_j$:
\begin{align}\label{chargegrid}
q(\mathbf{r}_j) &\equiv \sum\limits_s Z_s e \int\limits_{-\infty}^\infty \bar{g}(\mathbf{r}_j,v_\|) dv_\|  \\
&\approx \frac{1}{N_c} e \sum\limits_\beta n_{0\beta} Z_\beta S_3\left( \mathbf{R}_\beta - \mathbf{r}_j \right) w_\beta . \nonumber
\end{align}
The average number of particles of each species per grid cell is $N_c$, so that the total number of particles per species is $N_p = N_g N_c$. Finally, we can calculate $\phi_\mathbf{k}$ from the discrete Fourier transform of this quantity using equation \eqref{qneutralitysimple}:
\begin{equation}\label{phifromq}
\phi_\mathbf{k} = \frac{2 T_i}{n_{0i} e^2 k_\perp^2 \rho_i^2} \sum\limits_j e^{-i \mathbf{k}\cdot \mathbf{r}_j} q(\mathbf{r}_j) 
\end{equation}

Equations \eqref{gkcharacteristics}, \eqref{charcurves}, \eqref{fieldgrid2particle}, \eqref{fieldfourier}, \eqref{phifromq}, and \eqref{chargegrid} fully represent the $\delta f$-PIC algorithm as implemented with no approximation, and are combined to give equation \eqref{matrixeqn}.

\
\bibliographystyle{unsrt}
\bibliography{library}

\end{document}